\documentclass[showpacs,aps,graphicx,twocolumn]{revtex4}
\usepackage{graphicx}

\begin{document}

\title{Multipartite electronic entanglement purification with charge detection\footnote{Published in Phys. Lett. A 375 (2011) 396-400}}

\author{Yu-Bo Sheng$^{1}$, Fu-Guo Deng$^{2}$,   and Gui-Lu Long$^{1,3,4}$\footnote{
Corresponding author. Email address: gllong@tsinghua.edu.cn.}}
\address{$^1$ Department of Physics, Tsinghua University,
Beijing 100084  China\\
$^2$ Department of Physics,  Beijing Normal University, Beijing 100875, China\\
$^3$ Key Laboratory for Atomic and Molecular NanoSciences, Tsinghua
University,
Beijing 100084, China\\
$^4$ Tsinghua National Laboratory for Information Science and
Technology, Beijing 100084, China }

\date{\today }

\begin{abstract}
We present a multipartite entanglement purification scheme in a
Greenberger-Horne-Zeilinger state for electrons based on their spins
and their charges.  This scheme works for purification with two
steps, i.e., bit-flipping error correction and phase-error flip
error correction. By repeating these two steps, the parties in
quantum communication can get some high-fidelity multipartite
entangled electronic systems.
\\
\\
\textbf{Keywords:} Quantum physics, Entanglement purification,
Multipartite, Electrons, Charge detection, Decoherence
\end{abstract}
\pacs{03.67.Pp Quantum error correction and other methods for
protection against decoherence - 03.67.Hk Quantum communication}
\maketitle

\section{introduction}

Quantum entanglement is of vital importance in achieving tasks of
quantum information processing and transmission \cite{book}. Most of
the practical quantum computation and quantum communication tasks
require that the separated parties in distant locations share the
maximally entangled state
\cite{Ekert91,BBM92,rmp,LongLiu,CORE,lixh,teleportation,densecoding,densecoding2}.
Especially, multipartite entangled states have many important
applications in quantum  computation \cite{book} and quantum
communication, such as controlled teleportation
\cite{cteleportation,cteleportation2}, quantum secret sharing
\cite{QSS,QSS2,QSS3}, quantum state sharing
\cite{QSTS,QSTS2,QSTS3,QSTS4,QSTS5}, and so on. All these tasks
require multipartite entangled states to set up the quantum channel
between legitimate participants in quantum communication. However,
with local operations and classical communication, the two users in
quantum communication can not create entanglement. If they want to
share the entanglement separately, they have to create the entangled
states locally and transmit them in a quantum channel. In a
practical transmission, the channel noise cannot be avoided, which
will make a maximally entangled state become a mixed one. This will
decrease the fidelity of quantum teleportation or make quantum
communication insecure.

Entanglement purification  provides us a powerful tool to distil
high-fidelity entangled states from less entangled ones
\cite{Bennett1,Deutsch,Pan1,Simon,shengpra,shengpra2,shengpra3,Murao,Horodecki,Yong,chuan1,chuan2,shengepjd}.
The first entanglement purification protocol, which is based on the
quantum controlled-not (CNOT) logic operations, was proposed by
Bennett \emph{et al.} \cite{Bennett1} in 1996 for purifying a
two-qubit Werner state. So far, most of entanglement purification
protocols
\cite{Bennett1,Deutsch,Pan1,Simon,shengpra,shengpra2,shengpra3,chuan1,chuan2}
are focused on bipartite entangled quantum systems and there are
only several multipartite entanglement protocols, including
high-dimension entanglement protocols
\cite{Murao,Horodecki,Yong,shengepjd} as the latter is more
complicated than the former. In 1998, Murao \emph{et al}.
\cite{Murao} proposed the first multipartite entanglement
purification protocol with  CNOT logic operations for quantum
systems originally in a Greenberger-Horne-Zeilinger (GHZ) state.
This protocol was extended to high-dimensional multipartite quantum
systems in 2007 \cite{Yong}. They use some generalized XOR gates to
substitute the CNOT gates to fulfill their protocol.

Until now, most of the purification protocols are based on the
photons as they are manipulated easily. On the other hand,
conduction electrons can also be used to achieve quantum
communication and computation processes since Beenakker \emph{et
al.} \cite{beenakker} broke through the obstacle of the no-going
theorem \cite{nogo} in 2004. An electron acts as a qubit in both
charge degree of freedom and spin degree of freedom. If one measures
the charge degree of freedom of an electron quantum system, it will
leave its spin degree of freedom unaffected. By means of charge
detections \cite{cd}, Beenakker \emph{et al.} \cite{beenakker}
designed a protocol for a CONT gate between electronic qubits.
Moreover, people have constructed entangled spins \cite{paritybox},
achieved the entanglement concentration \cite{concentration},
prepared cluster states and designed a multipartite entanglement
analyzer \cite{cluster} with charge detections of electron quantum
systems. A bipartite entanglement purification protocol was also
presented in 2005 \cite{feng}, although their protocol is only used
to purify the Werner state, similar to the original entanglement
purification protocol by Bennett \emph{et al.} for photon pairs.
After each purification step, they have to add another bilateral
$\pi/2$ operations to recover the mixed state to Werner state in
order to perform the next purification step, which make its
efficiency low \cite{Deutsch}.

In this paper, we present a multipartite entanglement purification
for electrons in a Greenberger-Horne-Zeilinger state based on their
spins and their charges, resorting to charge detections and electron
polarizing beam splitters. In this scheme, charge detections play
the role of a parity check. The whole purification scheme can be
divided into two steps, i.e., bit-flip error purification and
phase-flip error purification. It dose not need to add another
operation to recover the mixed state to Werner state after each
purifying step. That is, it is easier than the only one entanglement
purification scheme for two-electron system \cite{feng}, and it will
have a practical application in solid quantum computation and
communication.

\section{Multipartite electronic entanglement purification with charge detections}

Before we start to explain our purification scheme, we first
introduce a basic element for entanglement purification scheme,
i.e., a parity-check gate. Parity-check gates can be used to
construct a Bell-state analyzer and a CNOT gate \cite{beenakker}. In
Fig.1, the polarizing beam splitter (PBS) is used to transmit an
electron in the spin-up state $|\uparrow\rangle$ and reflect an
electron in the spin-down state $|\downarrow\rangle$. The charge
detector (C) can distinguish the occupation number one from the
occupation number 0 and 2, but it cannot distinguish the electron
numbers between 0 and 2. That is, it can distinguish the occupation
number even or odd. Let us suppose that  two polarization qubits $a$
and $b$ initially in the states
\begin{eqnarray}
|\Psi_{a}\rangle=\alpha_{1}|\uparrow\rangle+\alpha_{2}|\downarrow\rangle
\end{eqnarray}
and
\begin{eqnarray}
|\Psi_{b}\rangle=\beta_{1}|\uparrow\rangle+\beta_{2}|\downarrow\rangle.
\end{eqnarray}
Here
\begin{eqnarray}
|\alpha_{1}|^{2}+|\alpha_{2}|^{2} &=& 1,\nonumber\\
|\beta_{1}|^{2}+|\beta_{2}|^{2} &=& 1.
\end{eqnarray}
These two qubits  are transmitted into the spatial modes $a_{1}$ and
$b_{1}$, respectively, and they interact with each other on the PBS.
The whole state of the two electrons will evolve to
\begin{eqnarray}
|\Psi_{T}\rangle_{ab} &=&
\alpha_{1}\beta_{1}|\uparrow\uparrow\rangle_{ab} +
\alpha_{1}\beta_{2}|\uparrow\downarrow\rangle_{ab} \nonumber\\
&+& \alpha_{2}\beta_{1}|\downarrow\uparrow\rangle_{ab} +
\beta_{2}\beta_{2}|\downarrow\downarrow\rangle_{ab}.
\end{eqnarray}
One observes immediately that the states  $|\uparrow\uparrow\rangle$
and $|\downarrow\downarrow\rangle$ will lead the charge detection to
have the charge occupation number $C=1$ as each electron passes
through a different path after the first PBS. The states
$|\uparrow\downarrow\rangle$ and $|\downarrow\uparrow\rangle$ will
lead the charge detection to $C=0$ and $C=2$, respectively. The
charge detection cannot distinguish 0 and 2, and  it will show the
same result, i.e., $C=0$ for simplicity. The states
$|\uparrow\uparrow\rangle$ and $|\downarrow\downarrow\rangle$ can be
distinguished from $|\uparrow\downarrow\rangle$ and
$|\downarrow\uparrow\rangle$ by the different outcomes of the charge
detection. So this device can be used to accomplish a parity check
on a two-electron system. That is, one can get
$\alpha_{1}\beta_{1}|\uparrow\uparrow\rangle+\beta_{2}\beta_{2}|\downarrow\downarrow\rangle$
from the outport $a_{2}b_{2}$ if $C=1$ and get
$\alpha_{1}\beta_{2}|\uparrow\downarrow\rangle
+\alpha_{2}\beta_{1}|\downarrow\uparrow\rangle$ if $C=0$.

\begin{figure}[!h]
\begin{center}
\includegraphics[width=8cm,angle=0]{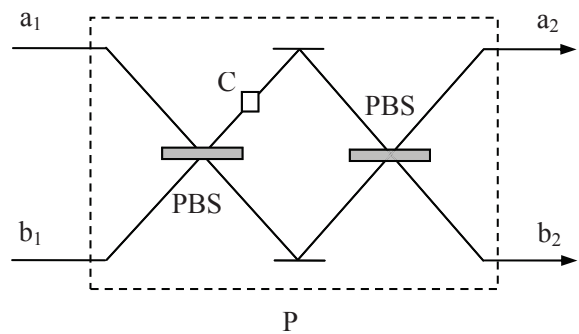}
\caption{ Schematic diagram showing the principle of a parity-check
gate. The parity-check gate (P) is composed of a  50:50 spin
polarizing beam splitters (PBS), a charge detection, and two
mirrors. The PBS is used to transmit an electron in the spin-up
state $|\uparrow\rangle$ and reflect an electron in the spin-down
state $|\downarrow\rangle$, respectively.}
\end{center}
\end{figure}

Now, let us detail how this entanglement purification scheme works
for multipartite electron systems. A multipartite
Greenberger-Horne-Zeilinger (GHZ) state for spin 1/2 systems can be
described as
\begin{eqnarray}
|\phi^{+}\rangle_s=\frac{1}{\sqrt{2}}(|\uparrow\uparrow\cdots\uparrow\rangle
+ |\downarrow\downarrow\cdots\downarrow\rangle).
\end{eqnarray}
We first take three-particle electron systems in GHZ states as an
example to show the principle of this multipartite entanglement
purification scheme and then extend to the case of $N$-particle
systems.

There are eight three-particle GHZ states, i.e.,
\begin{eqnarray}
|\Phi^{\pm}\rangle_{ABC}=\frac{1}{\sqrt{2}}(|\uparrow\uparrow\uparrow\rangle
\pm|\downarrow\downarrow\downarrow\rangle)_{ABC},\nonumber\\
|\Phi_{1}^{\pm}\rangle_{ABC}=\frac{1}{\sqrt{2}}(|\downarrow\uparrow\uparrow\rangle
\pm|\uparrow\downarrow\downarrow\rangle)_{ABC},\nonumber\\
|\Phi_{2}^{\pm}\rangle_{ABC}=\frac{1}{\sqrt{2}}(|\uparrow\downarrow\uparrow\rangle
\pm|\downarrow\uparrow\downarrow\rangle)_{ABC},\nonumber\\
|\Phi_{3}^{\pm}\rangle_{ABC}=\frac{1}{\sqrt{2}}(|\uparrow\uparrow\downarrow\rangle
\pm|\downarrow\downarrow\uparrow\rangle)_{ABC}.\label{GHZstate}
\end{eqnarray}
Here the subscripts A, B, and C represent the three electrons
belonging to the three parities, say Alice, Bob, and Charlie,
respectively. Initially, we suppose that the original GHZ state
transmitted is $|\Phi^{+}\rangle_{ABC}$. The noisy channel will
degrade the state and make the initial state be a mixed one. For
example, the state $|\Phi^{+}\rangle_{ABC}$ may become
$|\Phi_{1}^{+}\rangle_{ABC}$, say a bit-flip error, or become
$|\Phi^{-}\rangle_{ABC}$, say a phase-flip error. Sometimes both a
bit-flip error and a phase-flip error will take place such as
$|\Phi_{2}^{-}\rangle_{ABC}$. So the task of purifying
three-electron entangled systems can be divided into two step, i.e.
purifying the bit-flip error and the phase-flip error. The principle
of this multipartite entanglement purification for electron systems
is shown in Fig.2.

\begin{figure}[!h]
\begin{center}
\includegraphics[width=8cm,angle=0]{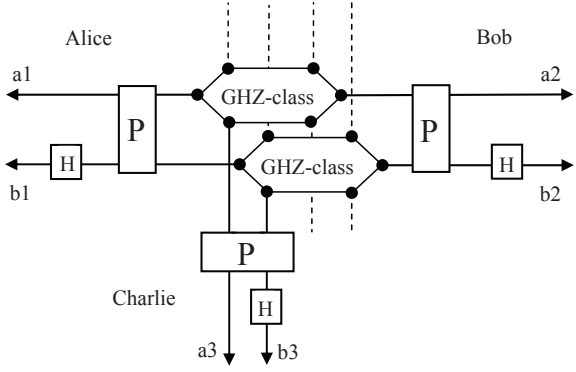}
\caption{ Schematic diagram showing the principle of multipartite
entanglement purification for electron systems. Each parties say
Alice, Bob and Charlie own the identical devices, respectively.  P
represents a parity check gate, which is used to discriminate the
parity for two electrons. A pair of GHZ states are transmitted to
each parties over a noisy channel. Each parties perform a parity
check, if both their parity are the same, they can achieve the
purification scheme.}
\end{center}
\end{figure}

\subsection{ Purification of bit-flip errors}

We first discuss the purification of bit-flip errors. Suppose that
an ensemble $\rho$ shared by Alice, Bob, and Charlie after the
transmission over a noisy channel is
\begin{eqnarray}
\rho=F|\Phi^{+}\rangle\langle\Phi^{+}|+(1-F)|\Phi_{1}^{+}\rangle\langle\Phi_{1}^{+}|.\label{ensemblerho}
\end{eqnarray}
Here $F>1/2$ is the fidelity of the state $|\Phi^{+}\rangle$ after
it is transmitted over a noisy channel and we suppose that a
bit-flip error takes place on the first electron with a probability
of $1-F$. In each step, each of the three parties will operate his
two electrons and each electron comes from one of the mixed state
$\rho$. For example, in Fig.2, we denote the two mixed states
$A_{1}B_{1}C_{1}$ and $A_{2}B_{2}C_{2}$. Then the electrons $A_{1}$
and $A_{2}$ belong to Alice, $B_{1}$ and $B_{2}$ belong to Bob, and
$C_{1}$ and $C_{2}$ belong to Charlie. From Eq.(\ref{ensemblerho}),
the state of the quantum system composed of the six electrons
$A_1B_1C_1A_2B_2C_2$ can be viewed as the mixture of four pure
states: it is in the state $|\Phi^{+}\rangle\otimes|\Phi^{+}\rangle$
with a probability of $F^{2}$, in the state
$|\Phi^{+}\rangle\otimes|\Phi^{+}_{1}\rangle$ and
$|\Phi^{+}_{1}\rangle\otimes|\Phi^{+}\rangle$ with an equal
probability of $F(1-F)$, and  in the state
$|\Phi^{+}_{1}\rangle\otimes|\Phi^{+}_{1}\rangle$ with a probability
of $(1-F)^{2}$. After the parity-check gates, these three parties
compare the parity of their electrons. They choose the cases that
all of them obtain the outcomes with the same parity, i.e., they all
obtain the even-parity ($C=1$) result or the odd-parity ($C=0$)
result.

We discuss the even-parity case first. The whole system is in a
mixed state by mixing two pure states, i.e.,
\begin{eqnarray}
|\phi\rangle=\frac{1}{\sqrt{2}}(|\uparrow\uparrow\uparrow\uparrow\uparrow\uparrow\rangle
+|\downarrow\downarrow\downarrow\downarrow\downarrow\downarrow\rangle)_{A_1B_1C_1A_2B_2C_2}\label{even-partiy1}
\end{eqnarray}
with a probability of $\frac{1}{2}F^{2}$ and
\begin{eqnarray}
|\phi_{1}\rangle=\frac{1}{\sqrt{2}}(|\downarrow\uparrow\uparrow\downarrow\uparrow\uparrow\rangle
+|\uparrow\downarrow\downarrow\uparrow\downarrow\downarrow\rangle)_{A_1B_1C_1A_2B_2C_2}\label{even-partiy2}
\end{eqnarray}
with a probability of $\frac{1}{2}(1-F)^2$. The cross-combinations
$|\Phi^{+}\rangle\otimes|\Phi^{+}_{1}\rangle$ and
$|\Phi^{+}_{1}\rangle\otimes|\Phi^{+}\rangle$ never lead  all the
three parties to have the same parity and can be eliminated
automatically. In the spatial modes $b_{1}b_{2}b_{3}$, Hadamard (H)
operations are performed on the electrons $A_{2}B_{2}C_{2}$, which
will lead to the transformation
\begin{eqnarray}
|\uparrow\rangle & \rightarrow & \frac{1}{\sqrt{2}}(|\uparrow\rangle+|\downarrow\rangle),\\
|\downarrow\rangle & \rightarrow &
\frac{1}{\sqrt{2}}(|\uparrow\rangle-|\downarrow\rangle).
\end{eqnarray}
After performing an $H$ operation on each of the electrons
$A_{2}B_{2}C_{2}$, Eq.(\ref{even-partiy1}) and
Eq.(\ref{even-partiy2}) become
\begin{eqnarray}
|\phi\rangle^{'}&=&\frac{1}{4}[|\uparrow\uparrow\uparrow\rangle(|\uparrow\rangle
+|\downarrow\rangle)^{\otimes3}  +
|\downarrow\downarrow\downarrow\rangle(|\uparrow\rangle
-|\downarrow\rangle)^{\otimes3}]\nonumber\\
&=&\frac{1}{4}[(\vert \uparrow\uparrow\uparrow \rangle +
|\downarrow\downarrow\downarrow\rangle)(\vert
\uparrow\uparrow\uparrow \rangle + \vert
\uparrow\downarrow\downarrow\rangle + \vert
\downarrow\uparrow\downarrow\rangle \nonumber\\
&& +\; \vert \downarrow\downarrow\uparrow \rangle)+(\vert
\uparrow\uparrow\uparrow \rangle -
|\downarrow\downarrow\downarrow\rangle)(\vert
\uparrow\uparrow\downarrow \rangle \nonumber\\
&& +\; \vert \uparrow\downarrow\uparrow\rangle + \vert
\downarrow\uparrow\uparrow\rangle + \vert
\downarrow\downarrow\downarrow \rangle)],\nonumber\\
|\phi_1\rangle^{'}&=&\frac{1}{4}[|\downarrow\uparrow\uparrow\rangle(|\uparrow\rangle
-|\downarrow\rangle)(|\uparrow\rangle
+ |\downarrow\rangle)^{\otimes2}\nonumber\\
&& + \; |\uparrow\downarrow\downarrow\rangle(|\uparrow\rangle
+|\downarrow\rangle)(|\uparrow\rangle-|\downarrow\rangle)^{\otimes2}]\nonumber\\
&=&\frac{1}{4}[(\vert \downarrow\uparrow\uparrow \rangle +
|\uparrow\downarrow\downarrow\rangle)(\vert \uparrow\uparrow\uparrow
\rangle + \vert \uparrow\downarrow\downarrow\rangle \nonumber\\
&& -\; \vert \downarrow\uparrow\downarrow\rangle - \vert
\downarrow\downarrow\uparrow \rangle)+(\vert
\downarrow\uparrow\uparrow \rangle -
|\uparrow\downarrow\downarrow\rangle)(\vert
\uparrow\uparrow\downarrow \rangle \nonumber\\
&& +\; \vert \uparrow\downarrow\uparrow\rangle - \vert
\downarrow\uparrow\uparrow\rangle - \vert
\downarrow\downarrow\downarrow \rangle)].
\end{eqnarray}
Finally, Alice, Bob, and Charlie measure the polarization states of
their electrons $A_2B_2C_2$ in the modes $b_{1}b_{2}b_{3}$. If they
obtain the outcomes $|\uparrow\uparrow\uparrow\rangle_{A_2B_2C_2}$,
$|\uparrow\downarrow\downarrow\rangle_{A_2B_2C_2}$,
$|\downarrow\uparrow\downarrow\rangle_{A_2B_2C_2}$ or
$|\downarrow\downarrow\uparrow\rangle_{A_2B_2C_2}$, they will obtain
the GHZ state $\frac{1}{\sqrt{2}}(\vert \uparrow\uparrow\uparrow
\rangle + |\downarrow\downarrow\downarrow\rangle)_{A_1B_1C_1}$ with
a probability of $\frac{1}{2}F^{2}$ and the GHZ state
$\frac{1}{\sqrt{2}}(\vert \downarrow\uparrow\uparrow \rangle +
|\uparrow\downarrow\downarrow\rangle)_{A_1B_1C_1}$ with a
probability of $\frac{1}{2}(1-F)^{2}$. That is, Alice, Bob, and
Charlie will get a new ensemble $\rho'$ with the fidelity of
$F'=\frac{F^{2}}{F^{2}+(1-F)^{2}}>F$ when $F>1/2$ by keeping the
electron systems $A_1B_1C_2$ if they only obtain one of the four
outcomes $\{|\uparrow\uparrow\uparrow\rangle_{A_2B_2C_2},
|\uparrow\downarrow\downarrow\rangle_{A_2B_2C_2},
|\downarrow\uparrow\downarrow\rangle_{A_2B_2C_2},
|\downarrow\downarrow\uparrow\rangle_{A_2B_2C_2}\}$. Certainly, they
will get the outcomes $|\uparrow\uparrow\downarrow\rangle$,
$|\uparrow\downarrow\uparrow\rangle$,
$|\downarrow\uparrow\uparrow\rangle$ or
$|\downarrow\downarrow\downarrow\rangle$. In this time, Alice, Bob,
and Charlie need only flip the relative phase of the electron system
$A_1B_1C_1$ and will obtain the ensemble $\rho'$.

On the other hand, if Alice, Bob, and Charlie all obtain an odd
parity, they will get the  state
\begin{eqnarray}
|\phi\rangle^o=\frac{1}{\sqrt{2}}(|\uparrow\uparrow\uparrow\downarrow\downarrow\downarrow\rangle
+|\downarrow\downarrow\downarrow\uparrow\uparrow\uparrow\rangle)_{A_1B_1C_1A_2B_2C_2}
\end{eqnarray}
with a probability of  $\frac{1}{2}F^{2}$ and the state
\begin{eqnarray}
|\phi_{1}\rangle^o=\frac{1}{\sqrt{2}}(|\downarrow\uparrow\uparrow\uparrow\downarrow\downarrow\rangle
+|\uparrow\downarrow\downarrow\downarrow\uparrow\uparrow\rangle)_{A_1B_1C_1A_2B_2C_2}
\end{eqnarray}
with a probability of  $\frac{1}{2}(1-F)^{2}$. Compared with
Eq.(\ref{even-partiy1}) and (\ref{even-partiy2}), Alice, Bob, and
Charlie only need to add a bit-flip operation on each of the three
qubits $A_{2}B_{2}C_{2}$ and they can get the same result as that
with the even parity.

By far, we have discussed the principle of the purification of the
bit-flip errors for three-electron systems. This method can also be
extended to purify the bit-flip errors in multipartite entangled
systems. For example, the initial state  of an $N$-electron quantum
system can be described as:
\begin{eqnarray}
|\Phi^+\rangle_{N}=\frac{1}{\sqrt{2}}(|\uparrow\uparrow\cdots
\uparrow\rangle+|\downarrow\downarrow\cdots
\downarrow\rangle)\label{stateN}.
\end{eqnarray}
Suppose a bit-flip error may take place in the first electron. After
transmission, each party makes a parity check on his two electrons
coming from two entangled quantum systems and they all choose the
same parity by classical communication, shown in Fig.2. Now, let us
suppose they all choose the even parity case and then the original
mixed state system becomes
\begin{eqnarray}
|\phi\rangle_{2N}=\frac{1}{\sqrt{2}}(|\uparrow\uparrow\uparrow\cdots\uparrow\rangle
+|\uparrow\uparrow\uparrow\cdots\uparrow\rangle)\label{correctN}
\end{eqnarray}
with a probability of  $\frac{1}{2}F^{2}$ and
\begin{eqnarray}
|\phi_{1}\rangle_{2N}=\frac{1}{\sqrt{2}}(|\downarrow\uparrow\cdots\uparrow\downarrow\uparrow\cdots\uparrow\rangle+
|\uparrow\downarrow\cdots\downarrow\uparrow\downarrow\cdots\uparrow\rangle)\label{errorN}
\end{eqnarray}
with a probability of  $\frac{1}{2}(1-F)^{2}$. After the H
operations on the $b_{1}b_{2}\cdots b_{N}$ modes,
Eq.(\ref{correctN}) and Eq.(\ref{errorN}) become
\begin{eqnarray}
|\phi\rangle^{'}_{2N}&=&\frac{1}{\sqrt{2}}(|\uparrow\uparrow\cdots\uparrow\rangle(\frac{1}{\sqrt{2}})^{\otimes
N}(|\uparrow\rangle
 + |\downarrow\rangle)^{\otimes N}\nonumber\\
&& +\;
|\downarrow\downarrow\cdots\downarrow\rangle(\frac{1}{\sqrt{2}})^{\otimes
N}(|\uparrow\rangle-|\downarrow\rangle)^{\otimes N}),\\
|\phi_{1}\rangle^{'}_{2N}&=&\frac{1}{\sqrt{2}}(|\downarrow\uparrow\cdots\uparrow\rangle(\frac{1}{\sqrt{2}})^{\otimes
N}(|\uparrow\rangle
 - |\downarrow\rangle)(|\uparrow\rangle\nonumber\\
 && +\; |\downarrow\rangle)^{\otimes(N-1)}
+|\uparrow\downarrow\cdots\downarrow\rangle(\frac{1}{\sqrt{2}})^{\otimes
N}(|\uparrow\rangle\nonumber\\
&& +\;
|\downarrow\rangle)(|\uparrow\rangle-|\downarrow\rangle)^{\otimes(N-1)}).
\end{eqnarray}
After the measurements on the electrons in the modes
$b_{1}b_{2}\cdots b_{N}$ with the basis
$Z$=$\{|\uparrow\rangle,|\downarrow\rangle\}$, the parties will
obtain a new ensemble $\rho''$ in which the  fidelity of the state
$|\Phi^{+}\rangle_{N}$  is $\frac{F^{2}}{F^{2}+(1-F)^{2}}$ if the
number of $|\downarrow\rangle$ is even. They will get the same
result with a phase-flip operation on each $N$-electron system kept
if the number of the outcomes $|\downarrow\rangle$ is odd.

We have fully described the principle of bit-flip error purification
on the first electron. If the bit-flip error takes place on  other
electrons, we can purify these errors in the same way and will get
the same result like those discussed above.

\subsection{ Purification of phase-flip errors}

Now we start to explain the principle of the phase-flip error
purification in the present scheme. Usually, during the
transmission, the relative phase between several entangled electrons
are sensitive to path length instabilities, which have to be kept
constant. This problem is analogous to the optical system for
quantum communications \cite{repeater1,phase}. Phase-flip errors
cannot be purified directly, but it can be converted into bit-flip
errors. For example,
 the eight GHZ states shown in Eq.(\ref{GHZstate})  can be transformed
into the following eight states by adding an $H$ operation on each
electron, i.e.,
\begin{eqnarray}
|\Psi^{+}\rangle=\frac{1}{2}(|\uparrow\uparrow\uparrow\rangle
+|\uparrow\downarrow\downarrow\rangle+|\downarrow\uparrow\downarrow\rangle+|\downarrow\downarrow\uparrow\rangle),\nonumber\\
|\Psi^{-}\rangle=\frac{1}{2}(|\uparrow\uparrow\downarrow\rangle
+|\uparrow\downarrow\uparrow\rangle+|\downarrow\uparrow\uparrow\rangle+|\downarrow\downarrow\downarrow\rangle),\nonumber\\
|\Psi_{1}^{+}\rangle=\frac{1}{2}(|\uparrow\uparrow\uparrow\rangle
+|\uparrow\downarrow\downarrow\rangle-|\downarrow\uparrow\downarrow\rangle-|\downarrow\downarrow\uparrow\rangle),\nonumber\\
|\Psi_{1}^{-}\rangle=\frac{1}{2}(|\uparrow\uparrow\downarrow\rangle
+|\uparrow\downarrow\uparrow\rangle-|\downarrow\uparrow\uparrow\rangle-|\downarrow\downarrow\downarrow\rangle),\nonumber\\
|\Psi_{2}^{+}\rangle=\frac{1}{2}(|\uparrow\uparrow\uparrow\rangle
-|\uparrow\downarrow\downarrow\rangle+|\downarrow\uparrow\downarrow\rangle-|\downarrow\downarrow\uparrow\rangle),\nonumber\\
|\Psi_{2}^{-}\rangle=\frac{1}{2}(|\uparrow\uparrow\downarrow\rangle
-|\uparrow\downarrow\uparrow\rangle+|\downarrow\uparrow\uparrow\rangle-|\downarrow\downarrow\downarrow\rangle),\nonumber\\
|\Psi_{3}^{+}\rangle=\frac{1}{2}(|\uparrow\uparrow\uparrow\rangle
-|\uparrow\downarrow\downarrow\rangle-|\downarrow\uparrow\downarrow\rangle+|\downarrow\downarrow\uparrow\rangle),\nonumber\\
|\Psi_{3}^{-}\rangle=\frac{1}{2}(|\uparrow\uparrow\downarrow\rangle
-|\uparrow\downarrow\uparrow\rangle-|\downarrow\uparrow\uparrow\rangle+|\downarrow\downarrow\downarrow\rangle).\label{phaseflipstate}
\end{eqnarray}
Suppose a phase-flip error may occur in the first electron and the
initial state after the electron systems are transmitted over a
noisy channel becomes a mixed state as follows
\begin{eqnarray}
\rho_p=F|\Phi^{+}\rangle\langle\Phi^{+}|+(1-F)|\Phi^{-}\rangle\langle\Phi^{-}|.\label{phaseerror1}
\end{eqnarray}
After the transformation by adding an $H$ operation on each
electron, Eq.(\ref{phaseerror1}) becomes
\begin{eqnarray}
\rho'_p=F|\Psi^{+}\rangle\langle\Psi^{+}|+(1-F)|\Psi^{-}\rangle\langle\Psi^{-}|.\label{phaseerror2}
\end{eqnarray}
It is interesting to find that in the state $|\Psi^{+}\rangle$, the
number of $|\downarrow\rangle$ in each items is even, but it is odd
in the state $|\Psi^{-}\rangle$. We also find that all the GHZ
states with the superscript $+$ in Eq.(\ref{phaseflipstate}) have
the even number of $|\downarrow\rangle$ but have the odd number of
$|\downarrow\rangle$ for $-$.

Now we detail the principle of the phase-flip error purification,
shown in  Fig.2. For two pairs $A_{1}B_{1}C_{1}$ and
$A_{2}B_{2}C_{2}$ picked out from the ensemble $\rho'_p$, their
state can be viewed as the mixture of four pure states:
$|\Psi^{+}\rangle \otimes |\Psi^{+}\rangle$, $|\Psi^{+}\rangle
\otimes |\Psi^{-}\rangle$, $|\Psi^{-}\rangle \otimes
|\Psi^{+}\rangle$, and $|\Psi^{-}\rangle \otimes |\Psi^{-}\rangle$
with the probabilities of $F^{2}$, $F(1-F)$, $F(1-F)$, and
$(1-F)^{2}$, respectively. Each party makes a parity-check
measurement on his two electrons with charge detection and then all
parties check their results by classical communication. They only
choose the case that all of them get the even parity and they
discard the other cases. In this way, the cross-combinations
$|\Psi^{+}\rangle \otimes |\Psi^{-}\rangle$ and $|\Psi^{-}\rangle
\otimes |\Psi^{+}\rangle$ are eliminated automatically and the
remaining items are
\begin{eqnarray}
|\varphi\rangle &=&
\frac{1}{2}(|\uparrow\uparrow\uparrow\uparrow\uparrow\uparrow\rangle
+|\uparrow\downarrow\downarrow\uparrow\downarrow\downarrow\rangle  +
 |\downarrow\uparrow\uparrow\downarrow\uparrow\uparrow\rangle
\nonumber\\
&&
+\;|\downarrow\downarrow\uparrow\downarrow\downarrow\uparrow\rangle)_{A_1B_1C_1A_2B_2C_2}
\label{evenparity1}
\end{eqnarray}
and
\begin{eqnarray}
|\varphi'\rangle &=&
\frac{1}{2}(|\uparrow\uparrow\downarrow\uparrow\uparrow\downarrow\rangle
+|\uparrow\downarrow\uparrow\uparrow\downarrow\uparrow\rangle  +
|\downarrow\uparrow\uparrow\downarrow\uparrow\uparrow\rangle
\nonumber\\
&&
+\;|\downarrow\downarrow\downarrow\downarrow\downarrow\downarrow\rangle)_{A_1B_1C_1A_2B_2C_2}.
\label{evenparity2}
\end{eqnarray}
with the probabilities of $F^{2}$ and $(1-F)^{2}$, respectively. In
order to get the three-electron entangled state $|\Psi^{+}\rangle$,
the parties first perform an $H$ operation on each of the three
electrons $A_{2}B_{2}C_{2}$, which will make $|\varphi\rangle$ and
$|\varphi'\rangle$ evolve as
\begin{eqnarray}
|\varphi\rangle &\rightarrow& \frac{1}{4\sqrt{2}}
|\uparrow\uparrow\uparrow\rangle_{A_2B_2C_2}(|\uparrow\uparrow\uparrow\rangle
+|\uparrow\uparrow\downarrow\rangle+|\uparrow\downarrow\uparrow\rangle+|\uparrow\downarrow\downarrow\rangle\nonumber\\
&& +
|\downarrow\uparrow\uparrow\rangle+|\downarrow\uparrow\downarrow\rangle
+|\downarrow\downarrow\uparrow\rangle+|\downarrow\downarrow\downarrow\rangle)_{A_1B_1C_1}\nonumber\\
&&  +
|\uparrow\downarrow\downarrow\rangle_{A_2B_2C_2}(|\uparrow\uparrow\uparrow\rangle
-|\uparrow\uparrow\downarrow\rangle-|\uparrow\downarrow\uparrow\rangle+|\uparrow\downarrow\downarrow\rangle\nonumber\\
&& +
|\downarrow\uparrow\uparrow\rangle-|\downarrow\uparrow\downarrow\rangle
-|\downarrow\downarrow\uparrow\rangle+|\downarrow\downarrow\downarrow\rangle)_{A_1B_1C_1}\nonumber\\
&&  +
|\downarrow\uparrow\downarrow\rangle_{A_2B_2C_2}(|\uparrow\uparrow\uparrow\rangle
-|\uparrow\uparrow\downarrow\rangle+|\uparrow\downarrow\uparrow\rangle-|\uparrow\downarrow\downarrow\rangle\nonumber\\
&& -
|\downarrow\uparrow\uparrow\rangle+|\downarrow\uparrow\downarrow\rangle
-|\downarrow\downarrow\uparrow\rangle+|\downarrow\downarrow\downarrow\rangle)_{A_1B_1C_1}\nonumber\\
&&  +
|\downarrow\downarrow\uparrow\rangle_{A_2B_2C_2}(|\uparrow\uparrow\uparrow\rangle
+|\uparrow\uparrow\downarrow\rangle-|\uparrow\downarrow\uparrow\rangle-|\uparrow\downarrow\downarrow\rangle\nonumber\\
&& -
|\downarrow\uparrow\uparrow\rangle-|\downarrow\uparrow\downarrow\rangle
+|\downarrow\downarrow\uparrow\rangle+|\downarrow\downarrow\downarrow\rangle)_{A_1B_1C_1}
\label{finalstate1},
\end{eqnarray}
\begin{eqnarray}
|\varphi\rangle' & \rightarrow &\frac{1}{4\sqrt{2}}
|\uparrow\uparrow\downarrow\rangle_{A_2B_2C_2}(|\uparrow\uparrow\uparrow\rangle
-|\uparrow\uparrow\downarrow\rangle+|\uparrow\downarrow\uparrow\rangle-|\uparrow\downarrow\downarrow\rangle\nonumber\\
&& +
|\downarrow\uparrow\uparrow\rangle-|\downarrow\uparrow\downarrow\rangle
+|\downarrow\downarrow\uparrow\rangle-|\downarrow\downarrow\downarrow\rangle)_{A_1B_1C_1}\nonumber\\
&&   +
|\uparrow\downarrow\uparrow\rangle_{A_2B_2C_2}(|\uparrow\uparrow\uparrow\rangle
+|\uparrow\uparrow\downarrow\rangle-|\uparrow\downarrow\uparrow\rangle-|\uparrow\downarrow\downarrow\rangle\nonumber\\
&& +
|\downarrow\uparrow\uparrow\rangle+|\downarrow\uparrow\downarrow\rangle
-|\downarrow\downarrow\uparrow\rangle-|\downarrow\downarrow\downarrow\rangle)_{A_1B_1C_1}\nonumber\\
&&   +
|\downarrow\uparrow\uparrow\rangle_{A_2B_2C_2}(|\uparrow\uparrow\uparrow\rangle
+|\uparrow\uparrow\downarrow\rangle+|\uparrow\downarrow\uparrow\rangle+|\uparrow\downarrow\downarrow\rangle\nonumber\\
&&-|\downarrow\uparrow\uparrow\rangle-|\downarrow\uparrow\downarrow\rangle
-|\downarrow\downarrow\uparrow\rangle-|\downarrow\downarrow\downarrow\rangle)_{A_1B_1C_1}\nonumber\\
&&  +
|\downarrow\downarrow\downarrow\rangle_{A_2B_2C_2}(|\uparrow\uparrow\uparrow\rangle
-|\uparrow\uparrow\downarrow\rangle-|\uparrow\downarrow\uparrow\rangle+|\uparrow\downarrow\downarrow\rangle\nonumber\\
&& -
|\downarrow\uparrow\uparrow\rangle+|\downarrow\uparrow\downarrow\rangle
+|\downarrow\downarrow\uparrow\rangle-|\downarrow\downarrow\downarrow\rangle)_{A_1B_1C_1}.
\label{finalstate2}
\end{eqnarray}
Then they measure their electrons $A_2$, $B_2$ and $C_2$ with the
basis $Z=\{|\uparrow\rangle,|\downarrow\rangle\}$.

From Eq.(\ref{finalstate1}) and Eq.(\ref{finalstate2}), if the
number of the outcome $|\downarrow\rangle$ in the measurements on
$A_{2}B_{2}C_{2}$ is even, i.e., $|\uparrow\uparrow\uparrow\rangle$,
$|\uparrow\downarrow\downarrow\rangle$,
$|\downarrow\uparrow\downarrow\rangle$, or
$|\downarrow\downarrow\uparrow\rangle$, the three parties can get
$|\psi^{+}\rangle$ with the fidelity of
$\frac{F^{2}}{F^{2}+(1-F)^{2}}$. Otherwise, if it is odd, the three
parties will get the state $|\psi^{-}\rangle$ with the fidelity of
$\frac{F^{2}}{F^{2}+(1-F)^{2}}$. The remaining  state of
$|\psi^{+}\rangle$ or $|\psi^{-}\rangle$ can be transformed into
$|\Phi^{+}\rangle$ or $|\Phi^{-}\rangle$ by adding another $H$
operation on each electron.

For $N$-particle quantum systems, if a phase-flip error takes place,
the parties can also purify the error with the same method mentioned
above. First,  phase-flip errors can be converted into  bit-flip
errors with an $H$ operation on each electron. Second, each parity
perform a parity check on his two electrons and all parties keep
their electron system with the same parity. Finally, they measure
the electrons in the lower modes with the  basis $z$ after all the
parties perform an $H$ operation on each of the electrons in the
lower modes. If the number of outcome $|\downarrow\rangle$ in the
measurements is even, the parties fulfill a probabilistic
purification of phase-flip errors. After they perform an $H$
operation on each electron kept, the parties can get a new ensemble
with the fidelity of $\frac{F^{2}}{F^{2}+(1-F)^{2}}$ which is larger
than the original one $F$ if $F>1/2$.

\section{Discussion and Summary}

So far, we have fully described our purification scheme for
multipartite electronic entangled states. We have explained the
entanglement purification principle for multipartite electron
systems with a special density matrices. That is, we have discussed
the principle of our entanglement purification protocol for
purifying the bit-flip error on the first particle and the
phase-flip error, shown in Eqs. (\ref{ensemblerho}) and
(\ref{phaseerror1}). In fact, in a practical environment, the mixed
state may be the Werner-type state, or more complicated. For
instance, a general three-particle mixed state can be written as:
\begin{eqnarray}
\rho_{g}=F|\Phi^{+}\rangle\langle\Phi^{+}|+F_{1}|\Phi^{-}\rangle\langle\Phi^{-}|
+ \cdots + F_{7}|\Phi_{3}^{-}\rangle\langle\Phi_{3}^{-}|,
\label{errorarbitrary}
\end{eqnarray}
where $F+F_{1}+\cdots+F_{7}=1$. In order to increase the fidelity
$F$, we should purify each unwanted item like $|\Phi^{-}\rangle$,
$|\Phi_{1}^{+}\rangle$, $\cdots$, and $|\Phi_{3}^{-}\rangle$. The
state $\rho_{g}$ contains both bit-flip errors and phase-flip
errors. Alice,  Bob, and Charlie can purify first the bit-flip error
on the first particle and then the bit-flip error on the second
particle, whose principle is similar to the case with only the
bit-flip error on the first particle. In this way, Alice, Bob, and
Charlie can purify the bit-flip errors on an arbitrary position, as
discussed in Ref. \cite{Murao}. By an H operation on each particle,
the phase-flip errors in Eq.(\ref{errorarbitrary}) can be
transferred into the bit-flip errors, which can be purified with the
similar way to the latter. That is to say, a general mixed state of
multipartite entangled electron systems can be purified by repeating
both the bit-flip-error purification  and the phase-flip-error
purification, similar to the original polarization entanglement
purification scheme for multi-particle Boson systems \cite{Murao}.

Charge detection has played a prominent role in constructing the
parity-check gate, and also it is the key element of  the present
purification protocol. It has been realized by means of point
contacts in a two-dimensional electron gas. For instance,
Ref.\cite{cd} used the effect of the electric field of the charge on
the conductance of an adjacent point contact to realize the charge
detection. Ref.\cite{experiment1}  reported their experimental
results that the current achievable time resolution for charge
detection is $\mu s$. Trauzettel \emph{et al.} also proposed a
realization of a charge parity meter which is based on two double
quantum dots alongside a quantum point contact \cite{parity}. Their
realization of such a device can be seen as a specific example of
the general class of mesoscopic quadratic quantum measurement
detectors which is investigated by Mao \emph{et al.} \cite{parity2}.

In summary,   we have proposed a multipartite entanglement
purification protocol for electron systems with the help of
parity-check gates. We first use the electronic polarizing beam
splitters and charge detections to construct the parity check gate
and then  detail the multipartite entanglement purification for
electron systems in GHZ state. The present scheme does not require
the controlled-not gate and it works for purification with two
steps, i.e., bit-flipping error correction and phase-error flip
error correction. By repeating these two steps, the  parties in
quantum communication can get some high-fidelity multipartite
entangled electronic systems. These features will make this scheme
have a practical application in solid quantum computation and
communication in the future.

\section*{ACKNOWLEDGEMENTS}

Y.B.S. is supported by China Postdoctoral Science Foundation under
Grant No. 20090460365 and China Postdoctoral Special Science
Foundation under Grant No. 201003132. G.L.L. is supported by the
National Natural Science Foundation of China under Grant No.
10874098 and the National Basic Research Program of China under
Grant No. 2009CB929402. F.G.D. is supported by the National Natural
Science Foundation of China under Grant No. 10974020 and A
Foundation for the Author of National Excellent Doctoral
Dissertation of PR China under Grant No. 200723.


\end{document}